# AN IMPROVED APRIORI ALGORITHM FOR ASSOCIATION RULES

Mohammed Al-Maolegi[1], Bassam Arkok[2]

Computer Science, Jordan University of Science and Technology, Irbid, Jordan

## ABSTRACT

*There are several mining algorithms of association rules. One of the most popular algorithms is Apriori that is used to extract frequent itemsets from large database and getting the association rule for discovering the knowledge. Based on this algorithm, this paper indicates the limitation of the original Apriori algorithm of wasting time for scanning the whole database searching on the frequent itemsets, and presents an improvement on Apriori by reducing that wasted time depending on scanning only some transactions. The paper shows by experimental results with several groups of transactions, and with several values of minimum support that applied on the original Apriori and our implemented improved Apriori that our improved Apriori reduces the time consumed by 67.38% in comparison with the original Apriori, and makes the Apriori algorithm more efficient and less time consuming.*

## 1. INTRODUCTION

With the progress of the technology of information and the need for extracting useful information of business people from dataset [7], data mining and its techniques is appeared to achieve the above goal. Data mining is the essential process of discovering hidden and interesting patterns from massive amount of data where data is stored in data warehouse, OLAP (on line analytical process), databases and other repositories of information [11]. This data may reach to more than terabytes. Data mining is also called (KDD) knowledge discovery in databases [3], and it includes an integration of techniques from many disciplines such as statistics, neural networks, database technology, machine learning and information retrieval, etc [6]. Interesting patterns are extracted at reasonable time by KDD's techniques [2]. KDD process has several steps, which are performed to extract patterns to user, such as data cleaning, data selection, data transformation, data pre-processing, data mining and pattern evaluation [4].

The architecture of data mining system has the following main components [6]: data warehouse, database or other repositories of information, a server that fetches the relevant data from repositories based on the user's request, knowledge base is used as guide of search according to defined constraint, data mining engine include set of essential modules, such as characterization, classification, clustering, association, regression and analysis of evolution. Pattern evaluation module that interacts with the modules of data mining to strive towards interested patterns. Finally, graphical user interfaces from through it the user can communicate with the data mining system and allow the user to interact.

## 2. ASSOCIATION RULE MINING

Association Mining is one of the most important data mining's functionalities and it is the most popular technique has been studied by researchers. Extracting association rules is the core of data mining [8]. It is mining for association rules in database of sales transactions between items which is important field of the research in dataset [6]. The benefits of these rules are detecting unknown relationships, producing results which can perform basis for decision making and prediction [8]. The discovery of association rules is divided into two phases [10, 5]: detection the frequent itemsets and generation of association rules. In the first phase, every set of items is called itemset, if they occurred together greater than the minimum support threshold [9], this itemset is called frequent itemset. Finding frequent itemsets is easy but costly so this phase is more important than second phase. In the second phase, it can generate many rules from one itemset as in form, if itemset $\{I_1, I_2, I_3\}$, its rules are $\{I_1 \rightarrow I_2, I_3\}$, $\{I_2 \rightarrow I_1, I_3\}$, $\{I_3 \rightarrow I_1, I_2\}$, $\{I_1, I_2 \rightarrow I_3\}$, $\{I_1, I_3 \rightarrow I_1\}$, $\{I_2, I_3 \rightarrow I_1\}$, number of those rules is $n^2$-1 where n = number of items. To validate the rule (e.g. $X \rightarrow Y$), where X and Y are items, based on confidence threshold which determine the ratio of the transactions which contain X and Y to the transactions A% which contain X, this means that A% of the transactions which contain X also contain Y. minimum support and confidence is defined by the user which represents constraint of the rules. So the support and confidence thresholds should be applied for all the rules to prune the rules which it values less than thresholds values. The problem that is addressed into association mining is finding the correlation among different items from large set of transactions efficiency [8].

The research of association rules is motivated by more applications such as telecommunication, banking, health care and manufacturing, etc.

## 3. RELATED WORK

Mining of frequent itemsets is an important phase in association mining which discovers frequent itemsets in transactions database. It is the core in many tasks of data mining that try to find interesting patterns from datasets, such as association rules, episodes, classifier, clustering and correlation, etc [2]. Many algorithms are proposed to find frequent itemsets, but all of them can be catalogued into two classes: candidate generation or pattern growth.

Apriori [5] is a representative the candidate generation approach. It generates length (k+1) candidate itemsets based on length (k) frequent itemsets. The frequency of itemsets is defined by counting their occurrence in transactions. FP-growth, is proposed by Han in 2000, represents pattern growth approach, it used specific data structure (FP-tree), FP-growth discover the frequent itemsets by finding all frequent in 1-itemsets into condition pattern base , the condition pattern base is constructed efficiently based on the link of node structure that association with FP-tree. FP-growth does not generate candidate itemsets explicitly.

## 4. APRIORI ALGORITHM

Apriori algorithm is easy to execute and very simple, is used to mine all frequent itemsets in database. The algorithm [2] makes many searches in database to find frequent itemsets where k-itemsets are used to generate k+1-itemsets. Each k-itemset must be greater than or equal to minimum support threshold to be frequency. Otherwise, it is called candidate itemsets. In the first, the algorithm scan database to find frequency of 1-itemsets that contains only one item by counting each item in database. The frequency of 1-itemsets is used to find the itemsets in 2-itemsets which in turn is used to find 3-itemsets and so on until there are not any more k-itemsets.

If an itemset is not frequent, any large subset from it is also non-frequent [1]; this condition prune from search space in database.

# 5. LIMITATIONS OF APRIORI ALGORITHM

Apriori algorithm suffers from some weakness in spite of being clear and simple. The main limitation is costly wasting of time to hold a vast number of candidate sets with much frequent itemsets, low minimum support or large itemsets. For example, if there are $10^4$ from frequent 1-itemsets, it need to generate more than $10^7$ candidates into 2-length which in turn they will be tested and accumulate [2]. Furthermore, to detect frequent pattern in size 100 (e.g.) v1, v2… v100, it have to generate $2^{100}$ candidate itemsets [1] that yield on costly and wasting of time of candidate generation. So, it will check for many sets from candidate itemsets, also it will scan database many times repeatedly for finding candidate itemsets. Apriori will be very low and inefficiency when memory capacity is limited with large number of transactions.

In this paper, we propose approach to reduce the time spent for searching in database transactions for frequent itemsets.

# 6. THE IMPROVED ALGORITHM OF APRIORI

This section will address the improved Apriori ideas, the improved Apriori, an example of the improved Apriori, the analysis and evaluation of the improved Apriori and the experiments.

## 6.1. The improved Apriori ideas

In the process of Apriori, the following definitions are needed:

Definition 1: Suppose T={$T_1$, $T_2$, … , $T_m$},(m≥1) is a set of transactions, $_{Ti}$= {$I_1$, $_{I2}$, … , $I_n$},(n≥1) is the set of items, and k-itemset = {$i_1$, $i_2$, … , $i_k$},(k≥1) is also the set of k items, and k-itemset ⊆ I.
Definition 2: Suppose σ (itemset), is the support count of itemset or the frequency of occurrence of an itemset in transactions.
Definition 3: Suppose $C_k$ is the candidate itemset of size k, and $L_k$ is the frequent itemset of size k.

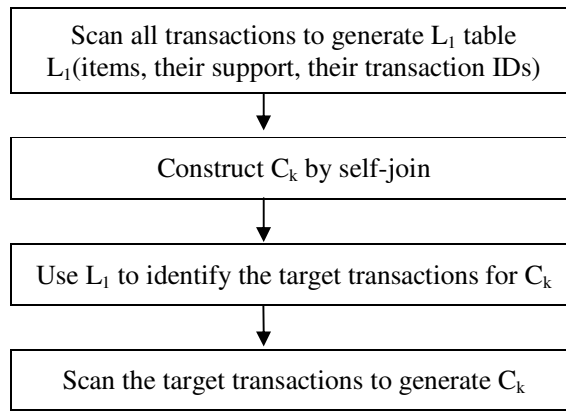

**Figure 1**: Steps for $C_k$ generation

In our proposed approach, we enhance the Apriori algorithm to reduce the time consuming for candidates itemset generation. We firstly scan all transactions to generate $L_1$ which contains the items, their support count and Transaction ID where the items are found. And then we use $L_1$ later as a helper to generate $L_2$, $L_3$ ... $L_k$. When we want to generate $C_2$, we make a self-join $L_1 * L_1$ to construct 2-itemset C (x, y), where x and y are the items of $C_2$. Before scanning all transaction records to count the support count of each candidate, use $L_1$ to get the transaction IDs of the minimum support count between x and y, and thus scan for $C_2$ only in these specific transactions. The same thing for $C_3$, construct 3-itemset C (x, y, z), where x, y and z are the items of $C_3$ and use $L_1$ to get the transaction IDs of the minimum support count between x, y and z, then scan for $C_3$ only in these specific transactions and repeat these steps until no new frequent itemsets are identified. The whole process is shown in the Figure 1.

## 6.2. The improved Apriori

The improvement of algorithm can be described as follows:

//Generate items, items support, their transaction ID
(1) $L_1$ = find_frequent_1_itemsets (T);
(2) For (k = 2; $L_{k-1} \neq \Phi$; k++) {
//Generate the Ck from the $L_{K-1}$
(3) $C_k$ = candidates generated from $L_{k-1}$;
//get the item $I_w$ with minimum support in $C_k$ using $L_1$,(1≤w≤k).
(4) x = Get _item_min_sup($C_k$, $L_1$);
// get the target transaction IDs that contain item x.
(5) Tgt = get_Transaction_ID(x);
(6) For each transaction t in Tgt Do
(7) Increment the count of all items in $C_k$ that are found in Tgt;
(8) $L_k$= items in $C_k \geq$ min_support;
(9) End;
(10) }

## 6.3. An example of the improved Apriori

Suppose we have transaction set D has 9 transactions, and the minimum support = 3. The transaction set is shown in Table.1.

**Table 1:** The transactions

| T_ID | Items |
|------|-------|
| $T_1$ | $I_1, I_2, I_5$ |
| $T_2$ | $I_2, I_4$ |
| $T_3$ | $I_2, I_4$ |
| $T_4$ | $I_1, I_2, I_4$ |
| $T_5$ | $I_1, I_3$ |
| $T_6$ | $I_2, I_3$ |
| $T_7$ | $I_1, I_3$ |
| $T_8$ | $I_1, I_2, I_3, I_5$ |
| $T_9$ | $I_1, I_2, I_3$ |

**Table 2:** The candidate 1-itemset

| Items | support | |
|-------|---------|---|
| $I_1$ | 6 | |
| $I_2$ | 7 | |
| $I_3$ | 5 | |
| $I_4$ | 3 | |
| $I_5$ | 2 | deleted |

firstly, scan all transactions to get frequent 1-itemset $l_1$ which contains the items and their support count and the transactions ids that contain these items, and then eliminate the candidates that are infrequent or their support are less than the min_sup. The frequent 1-itemset is shown in table 3.

**Table 3:** Frequent 1_itemset

| Items | support | T_IDs | |
|-------|---------|-------|---|
| $I_1$ | 6 | $T_1, T_4, T_5, T_7, T_8, T_9$ | |
| $I_2$ | 7 | $T_1, T_2, T_3, T_4, T_6, T_8, T_9$ | |
| $I_3$ | 5 | $T_5, T_6, T_7, T_8, T_9$ | |
| $I_4$ | 3 | $T_2, T_3, T_4$ | |
| $I_5$ | 2 | $T_1, T_8$ | deleted |

The next step is to generate candidate 2-itemset from $L_1$. To get support count for every itemset, split each itemset in 2-itemset into two elements then use l1 table to determine the transactions where you can find the itemset in, rather than searching for them in all transactions. for example, let's take the first item in table.4 ($I_1, I_2$), in the original Apriori we scan all 9 transactions to find the item ($I_1, I_2$); but in our proposed improved algorithm we will split the item ($I_1, I_2$) into $I_1$ and $I_2$ and get the minimum support between them using $L_1$, here i1 has the smallest minimum support. After that we search for itemset ($I_1, I_2$) only in the transactions $T_1, T_4, T_5, T_7, T_8$ and $T_9$.

**Table 4:** Frequent 2_itemset

| Items | support | Min | Found in | |
|-------|---------|-----|----------|---|
| $I_1, I_2$ | 4 | $I_1$ | $T_1, T_4, T_5, T_7, T_8, T_9$ | |
| $I_1, I_3$ | 4 | $I_3$ | $T_5, T_6, T_7, T_8, T_9$ | |
| $I_1, I_4$ | 1 | $I_4$ | $T_2, T_3, T_4$ | deleted |
| $I_2, I_3$ | 3 | $I_3$ | $T_5, T_6, T_7, T_8, T_9$ | |
| $I_2, I_4$ | 3 | $I_4$ | $T_2, T_3, T_4$ | |
| $I_3, I_4$ | 0 | $I_4$ | $T_2, T_3, T_4$ | deleted |

The same thing to generate 3-itemset depending on $L_1$ table, as it is shown in table 5.

**Table 5:** Frequent 3-itemset

| Items | support | Min | Found in | |
|-------|---------|-----|----------|---|
| I1, I2 , I3 | 2 | $I_3$ | $T_5, T_6, T_7, T_8, T_9$ | deleted |
| I1, I2 , I4 | 1 | $I_4$ | $T_2, T_3, T_4$ | deleted |
| I1, I3 , I4 | 0 | $I_4$ | $T_2, T_3, T_4$ | deleted |
| I2, I3 , I4 | 0 | $I_4$ | $T_2, T_3, T_4$ | deleted |

For a given frequent itemset $L_k$, find all non-empty subsets that satisfy the minimum confidence, and then generate all candidate association rules.

In the previous example, if we count the number of scanned transactions to get (1, 2, 3)-itemset using the original Apriori and our improved Apriori, we will observe the obvious difference between number of scanned transactions with our improved Apriori and the original Apriori. From the table 6, number of transactions in1-itemset is the same in both of sides, and whenever the k of k-itemset increase, the gap between our improved Apriori and the original Apriori increase from view of time consumed, and hence this will reduce the time consumed to generate candidate support count.

**Table 6:** Number of transactions scanned Experiments

|  | **Original Apriori** | **Our improved Apriori** |
|---|---|---|
| **1-itemset** | 45 | 45 |
| **2-itemset** | 54 | 25 |
| **3-itemset** | 36 | 14 |
| **sum** | **135** | **84** |

We developed an implementation for original Apriori and our improved Apriori, and we collect 5 different groups of transactions as the follow:

- T1: 555 transactions.
- T2: 900 transactions.
- T3: 1230 transactions.
- T4: 2360 transactions.
- T5: 3000 transactions.

The first experiment compares the time consumed of original Apriori, and our improved algorithm by applying the five groups of transactions in the implementation. The result is shown in Figure 2.

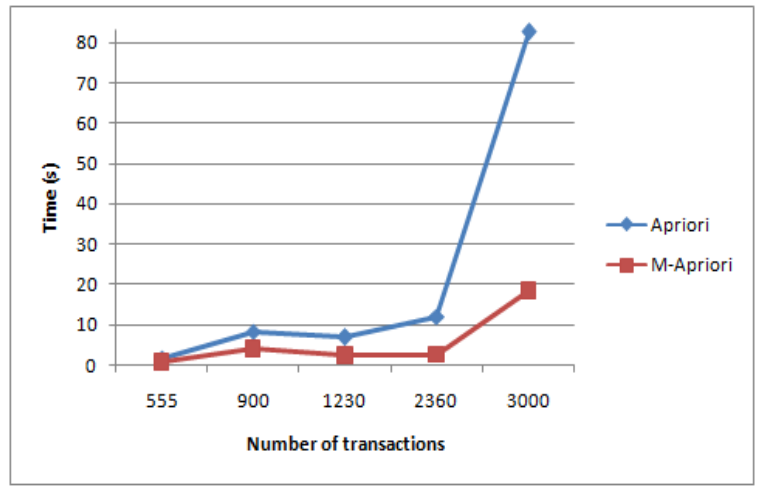

**Figure 2**: Time consuming comparison for different groups of transactions

The second experiment compares the time consumed of original Apriori, and our proposed algorithm by applying the one group of transactions through various values for minimum support in the implementation. The result is shown in Figure 3.

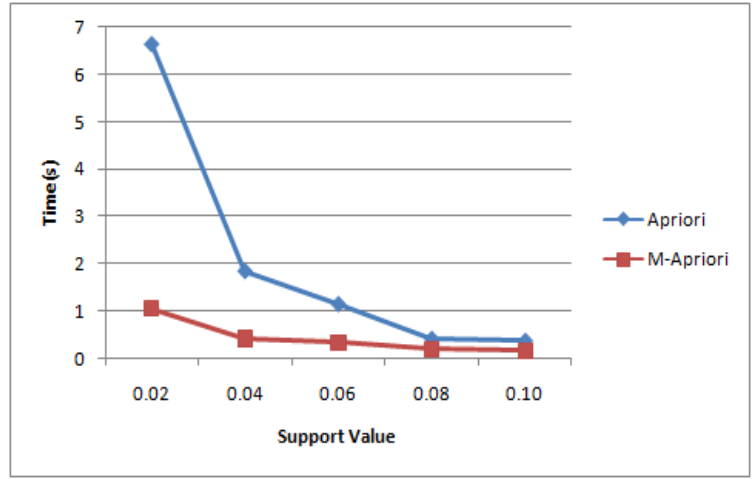

**Figure 3**: Time consuming comparison for different values of minimum support

### 6.4. The analysis and evaluation of the improved Apriori

As we observe in figure 2, that the time consuming in improved Apriori in each group of transactions is less than it in the original Apriori, and the difference increases more and more as the number of transactions increases.

Table 7 shows that the improved Apriori reduce the time consuming by 61.88% from the original Apriori in the first group of transactions T1, and by 77.80% in T5. As the number of transactions increase the rate is increased also. The average of reducing time rate in the improved Apriori is 67.38%.

**Table 7:** THE time reducing rate of improved Apriori on the original Apriori according to the number of transactions

| T | Original Apriori (S) | Improved Apriori (S) | Time reducing rate (%) |
|-------|----------------------|----------------------|------------------------|
| $T_1$ | 1.776 | 0.677 | 61.88% |
| $T_2$ | 8.221 | 4.002 | 51.32% |
| $T_3$ | 6.871 | 2.304 | 66.47% |
| $T_4$ | 11.940 | 2.458 | 79.41% |
| $T_5$ | 82.558 | 18.331 | 77.80% |

As we observe in figure 3, that the time consuming in improved Apriori in each value of minimum support is less than it in the original Apriori, and the difference increases more and more as the value of minimum support decreases.

Table 8 shows that the improved Apriori reduce the time consuming by 84.09% from the original Apriori where the minimum support is 0.02, and by 56.02% in 0.10. As the value of minimum support increase the rate is decreased also. The average of reducing time rate in the improved Apriori is 68.39%.

**Table 8:** The time reducing rate of improved Apriori on the original Apriori according to the value of minimum support

| Min_Sup | Original Apriori (S) | Improved Apriori (S) | Time reducing rate (%) |
|---------|----------------------|----------------------|------------------------|
| 0.02 | 6.638 | 1.056 | 84.09% |
| 0.04 | 1.855 | 0.422 | 77.25% |
| 0.06 | 1.158 | 0.330 | 71.50% |
| 0.08 | 0.424 | 0.199 | 53.07% |
| 0.10 | 0.382 | 0.168 | 56.02% |

## 7. CONCLUSION

In this paper, an improved Apriori is proposed through reducing the time consumed in transactions scanning for candidate itemsets by reducing the number of transactions to be scanned. Whenever the k of k-itemset increases, the gap between our improved Apriori and the original Apriori increases from view of time consumed, and whenever the value of minimum support increases, the gap between our improved Apriori and the original Apriori decreases from view of time consumed. The time consumed to generate candidate support count in our improved Apriori is less than the time consumed in the original Apriori; our improved Apriori reduces the time consuming by 67.38%. As this is proved and validated by the experiments and obvious in figure 2, figure 3, table 7 and table 8.

## ACKNOWLEDGEMENTS

We would like to thank all academic staff in our university for supporting us in each research projects specially this one.

**Authors**

Mohammed Al-Maolegi Obtained his Master degree in computer science from Jordan University of Science and Technology University (Jordan) in 2014. He received his B.Sc. in computer information system from Mutah University (Jordan) in 2010. His research interests include: softw are engineering, software metrics, data mining and wireless sensor networks. 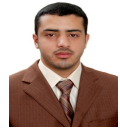

Bassam Arkok Obtained his Master degree in computer science from Jordan University of Science and Technology University (Jordan) in 2014. He received his B.Sc. in computer science from Alhodidah University (Yemen). His research interests include: software engineering, software metrics, data mining and wireless sensor networks.